\documentclass{article}
\usepackage{amsfonts, amsmath,graphicx,epstopdf,url,bm}
\title{Stern-Gerlach surfing in laser wakefield accelerators}
\date{\today}
\author{S. P. Flood\thanks{Department of Physics, Lancaster University, Lancaster, UK and Cockcroft Institute, Daresbury, UK} \and D. A. Burton\footnotemark[1]}
\begin{document}
\maketitle
\begin{abstract}

We investigate the effects of a Stern-Gerlach-type addition to the Lorentz force on electrons in a laser wakefield accelerator. The Stern-Gerlach-type terms are found to generate a family of trajectories describing electrons that `surf' along the plasma density wave driven by a laser pulse. Such trajectories could lead to an increase in the size of an electron bunch, which may have implications for attempts to exploit such bunches in future free electron lasers.

\end{abstract}

\section{Introduction}

The modelling of charged particles in electromagnetic fields has been an area of great interest to accelerator physicists for decades, using a number of simplifications for computational ease. Increasingly large fields such as those that will be produced at ELI \cite{ELI} and HiPER \cite{HiPER} now mean that the validity of these simplifications must be considered more closely.

The classical radiation reaction force has long been considered as an averaged effect for sufficiently low electromagnetic fields in accelerator physics and more recently the Landau-Lifshitz equation has been used to model the motion of such particles in detail \cite{berezhiani2004radiation,hazeltine2004radiation,tamburini2011radiation}. 
The Landau-Lifshitz equation is however only valid for sufficiently weak and slowly varying background fields \cite{BurtonNobleReview}.
The increasingly large electromagnetic fields expected to be deployed at experiments such as ELI and HiPER may be strong enough to render the Landau-Lifshitz equation invalid and this has led to recent interest in alternative models \cite{noblekravetsjaroszynski2013},  computational simplifications \cite{Vranicetal2015} and extensive consideration of quantum effects (for a recent review, see \cite{BurtonNobleReview}). Thus, calculation of the effective trajectories from first principles proves problematic, due to the various candidate effective theories. Recently \cite{Vranicetal2015}, various radiation reaction models have been been included in PIC codes via the Osiris framework, which is commonly used in the laser wakefield context.

While the effects of the increasingly large electromagnetic field {strengths} have been studied, the effects of large field \emph{gradients}, created in laser wakefield accelerators, have received very little attention.
In contexts where the field gradient is considered to be non-trivial the radiation reaction contributions are included via the Landau-Lifshitz equation, though the effects of spin are often neglected despite being of comparable size \cite{Piazza}.
The Stern-Gerlach force, caused by the coupling of the spin of a charged particle with the background electromagnetic field, was first observed  in the splitting of a beam of silver atoms by an inhomogeneous magnetic field.
In high field-gradient systems, such as those created in ultra-intense laser-driven plasma density waves, we suggest that the Stern-Gerlach force is not only non-trivial but in some circumstances may be \emph{more important} than the radiation reaction force. These Stern-Gerlach effects may lead to behaviour that does not appear to have been adequately addressed in the context of laser wakefield accelerators: the purpose of this article is to illustrate the significance of the Stern-Gerlach forces in a simple model of such an accelerator.

Modelling quantum mechanical electrons as covariant classical spinning particles has been well-studied. There have been various approaches from the work of Frenkel \cite{frenkel1926elektrodynamik} and Thomas \cite{thomas1927kinematics} in the 1920s through the work of Nakano \cite{nakano1956relativistic}, Tulczyjew \cite{tulczyjew1959motion}, Dixon \cite{dixon1964,dixon1965,dixon1967}, Corben \cite{corben1961spin1,corben1961spin2}, Suttorp and de Groot \cite{suttorpdegroot,de1972foundations} and Ellis \cite{ellis1975motion} in the 1950-70s. 
The equations of motion and the validity of the auxiliary condition on the spin and the momentum necessary to close the system have still received discussion in recent years \cite{Costaetal2012}.
The approaches used to derive these equations are varied, and for completeness we include a new method using de Rham currents and distributional methods in Appendix \ref{app:derivation}. 

The systems in which the Stern-Gerlach force is most prominent are those with a high electromagnetic field gradient.  Chapter 2 considers the implications of the coupling between the spin of a classical electron and the rapidly varying electromagnetic field produced by a laser-driven plasma wave.  Sufficiently short, high-intensity laser pulses can form longitudinal waves within the electron density of a plasma. These density waves propagate with speed comparable to the group speed of the laser pulse. Not all plasma electrons form this wave, however; some of the electrons are caught up in the wave and accelerated by its high fields. The wave eventually collapses as these electrons damp the wave (the wave `breaks'). The extremely high electric field gradient of a plasma wave near wavebreaking provides an excellent theoretical testing ground for the effects of Stern-Gerlach-type contributions to the trajectory of a test electron. 

In what follows, the equations of motion for a spinning electron in such a density wave are found to have a particular solution which does not exist for a particle without spin - trajectories corresponding to electrons `surfing' orthogonal to the wave vector in the frame of the wave.  The perturbations around a `surfing' trajectory are found to be linearly unstable for the vast majority of the parameter space. Since the family of new trajectories found in Section 2 correspond to electrons travelling orthogonal to the motion of the plasma electrons and are unstable, the electrons following such trajectories could cause undesirable properties for effective bunching of electrons in laser wakefield accelerators. 

These `surfing' trajectories exist only for a particle with non-zero spin in a background field with non-zero gradient. Furthermore, the electrons are non-accelerating and therefore the radiation reaction forces are expected to be negligible. Clearly, the spin-field coupling is much more significant than radiation reaction in the present context.


{~}

We use Heaviside-Lorentz units with the speed of light $c=1$ (except at the end of Section \ref{stabilitysection} for the sake of clarity) and we assume that the effects of spacetime curvature are negligible so that the spacetime metric is simply the Minkowski metric $g_{ab} = \text{diag}\{-1,1,1,1\}$. Lower case Latin indices run over 0, 1, 2, 3.

\section{Effects of Stern-Gerlach-type forces on a classical charged particle}

\subsection{Preliminaries}

The equations of motion that govern a classical particle with worldline $C : \tau \mapsto x^a=C^a(\tau)$, charge $q$, momentum $P^{a}$ and spin $S_{ab}$ in a background electromagnetic field described by the tensor components $F_{ab}$ are 
\begin{align}
&\frac{d}{d\tau}\left(P^a + \frac{{F}^{ab}\Sigma_{bc} {P}^c}{\dot{C}^dP_d}\right)= -q{F}^{ab}{\dot{C}_b}-\frac{1}{2}\Sigma^{bc}\partial^a{F}_{bc},\label{myeqgeneral1}\\
&\frac{d}{d\tau}S^{ab}= -\dot{C}^a\left(P^b + \frac{{F}^{bc}\Sigma_{cd} {P}^d}{\dot{C}^eP_e}\right)+\dot{C}^b\left(P^a + \frac{{F}^{ac}\Sigma_{cd} {P}^d}{\dot{C}^eP_e}\right) +{F}^{bc}\Sigma^{~a}_{c} -{F}^{ac}\Sigma^{~b}_{c},\label{myeqgeneral2}
\end{align}
where $\tau$ is the proper time of the particle, $\dot{C}^a = \frac{d}{d\tau}C^a(\tau)$ is the 4-velocity of the particle and $\Sigma_{ab}$ is the electromagnetic dipole tensor (see \cite{suttorpdegroot}, or for a new derivation of these equations, see Appendix A). However, this is not a complete system; an additional condition is required. There are a number of possible conditions, though two of the most commonly used are the Frenkel condition \cite{frenkel1926elektrodynamik}
\begin{align}
\dot{C}^aS_{ab} = 0\label{Frenkelcond}
\end{align}
and the Nakano-Tulczyjew \cite{nakano1956relativistic, tulczyjew1959motion} condition 
\begin{align}
P^aS_{ab} = 0.\label{Tulczcond}
\end{align}
The Frenkel condition, whilst being simple and intuitive, is considered by some to be unphysical since it yields helical solutions in field-free systems (sometimes called Zittebewegung) \cite{corben1961spin1,corben1961spin2}, though others argue against this unphysicality \cite{Costaetal2012}. 
We initially adopt the Nakano-Tulczyjew condition \eqref{Tulczcond}, which has already been abundantly studied \cite{dixon1964,dixon1965,dixon1967,suttorpdegroot} though we subsequently show that in fact the conditions \eqref{Frenkelcond} and \eqref{Tulczcond} are \emph{equivalent} to first order in $S_{ab}$.

A particle with spin has a magnetic dipole moment related to the spin by the gyromagnetic ratio $\frac{{g}q}{2M_0}$, where $M_0$ is the particle's rest mass and ${g}$ is the $g$-factor of the particle.  Furthermore, a particle with a zero electric dipole moment and a non-zero magnetic dipole moment is characterised by $\Sigma_{ab} = \frac{{g}q}{2M_0}S_{ab}$\label{sym:gnumber}\label{sym:restmassM0}. Thus the equations of motion, together with the condition \eqref{Tulczcond}, are 
\begin{align}
&\frac{d}{d\tau}P^a= -q{F}^{ab}{\dot{C}_b}-\frac{{g}q}{4M_0}S^{bc}\partial^a{F}_{bc},\label{eqofmotionPa}\\
&\frac{d}{d\tau}S^{ab}= -\dot{C}^aP^b+\dot{C}^bP^a + \frac{{g}q}{2M_0}{F}^{bc}S^{~a}_{c} -\frac{{g}q}{2M_0}{F}^{ac}S^{~b}_{c},
\end{align}
where the 4-momentum $P^a$ satisfies the condition
\begin{align}
P_b = -P^a\dot{C}_a\dot{C}_b - \frac{\left(S_{ab}+S_{ad}\dot{C}^d\dot{C}_b\right)}{P^e\dot{C}_e}\left(qF^{ac}\dot{C}_c+\frac{{g}q}{4M_0}\partial^aF_{cd}S^{cd}+\frac{{g}q}{2M_0}P^cF_{c}^{~a}\right),\label{Pfulleqdef}
\end{align}
found by differentiation of the Nakano-Tulczyjew condition \eqref{Tulczcond} with respect to $\tau$. Note that the first term on the right-hand side of \eqref{eqofmotionPa} is the standard Lorentz force on a charged particle and the second term, the coupling of the spin and the gradient of the electromagnetic field, is a Stern-Gerlach-type contribution.

A classical electron has $g$-factor equal to 2, charge $q=q_{\text{e}}=-e$ (where $e$ is the elementary charge) and rest mass $M_0 = m_{\text{e}}$ giving the system of equations
\begin{align}
&\frac{d}{d\tau}P^a= -q_{\text{e}}{F}^{ab}{\dot{C}_b}-\frac{q_{\text{e}}}{2m_{\text{e}}}S^{bc}\partial^a{F}_{bc},\label{syseqelectron1}\\
&\frac{d}{d\tau}S^{ab}= -\dot{C}^aP^b+\dot{C}^bP^a +\frac{q_{\text{e}}}{m_{\text{e}}}{F}^{bc}S^{~a}_{c} -\frac{q_{\text{e}}}{m_{\text{e}}}{F}^{ac}S^{~b}_{c},\label{syseqelectron2}\\
&P^aS_{ab}=0\label{syseqelectron3},\\
&P^a = -P^b\dot{C}_b\dot{C}^a - \frac{\left(S^{~a}_{b}+S_{bd}\dot{C}^d\dot{C}^a\right)}{P^e\dot{C}_e}\left(q_{\text{e}}F^{bc}\dot{C}_c+\frac{q_{\text{e}}}{m_{\text{e}}}\partial^bF_{cd}S^{cd}+\frac{q_{\text{e}}}{m_{\text{e}}}P^cF_{c}^{~b}\right)\label{syseqelectron4}.
\end{align}
Integration of \eqref{syseqelectron1}-\eqref{syseqelectron4} is far from straightforward due to constraints \eqref{syseqelectron3}, \eqref{syseqelectron4} and it is useful to reduce \eqref{syseqelectron1}-\eqref{syseqelectron4} to a model that captures the essential physics that we wish to explore.  In order to simplify the system, we choose to linearise \eqref{syseqelectron1}-\eqref{syseqelectron4} in the spin\footnote{For a system of equations linearised in $F_{ab}$, see \cite{suttorpdegroot}.} $S_{ab}$. Firstly, note that linearising the momentum condition \eqref{syseqelectron4} results in the straightforward expression
\begin{align}
P^a = \left(m_{\text{e}} + \frac{q_{\text{e}}}{2m_{\text{e}}}S^{bc}F_{bc}\right)\dot{C}^a,\label{PSexplicitansatz}
\end{align}
for the 4-momentum and hence the Frenkel condition \eqref{Frenkelcond} and the Nakano-Tulczyjew condition \eqref{Tulczcond} are equivalent to first order in $S^{ab}$. The linearised system of equations is
\begin{align}
&\frac{d}{d\tau}\dot{C}^a =  -\left(1 - \frac{q_{\text{e}}}{2m_{\text{e}}^2}S^{bc}F_{bc}\right)\frac{q_{\text{e}}}{m_{\text{e}}}{F}^{ab}{\dot{C}_b}-\frac{q_{\text{e}}}{2m_{\text{e}}^2}S^{bc}\left(\Pi^\perp_{\dot{C}}\right)^{ad}\partial_dF_{bc}\label{eqmotlinS1},\\
&\frac{d}{d\tau}S^{ab}= \frac{q_{\text{e}}}{m_{\text{e}}}\left({F}^{bc}S^{~a}_{c} -{F}^{ac}S^{~b}_{c}\right)\label{eqmotlinS2},\\
&\dot{C}^aS_{ab}=0,\label{Frenkelcond2}
\end{align}
along with \eqref{PSexplicitansatz}, where
%
$(\Pi^\perp_{\dot{C}})^{ab} = g^{ab} + \dot{C}^a\dot{C}^b$.
%
Notably, the effects of the third term (the Stern-Gerlach-type term) on the right-hand side of \eqref{eqmotlinS1} are most apparent in a system with a high field gradient. Although we can demand that \eqref{Frenkelcond2} is satisfied at a particular instant in proper time $\tau$, it is only satisfied to second order in $S^{ab}$ at other times.  Henceforth we adopt \eqref{eqmotlinS1}, \eqref{eqmotlinS2} as the system of equations for a classical electron with spin subject to $\dot{C}^aS_{ab}|_{\tau=0}=0$ in order to readily demonstrate the effects of the Stern-Gerlach term (the final term on the right-hand side) of \eqref{eqmotlinS1}.

\subsection{Effects of the Stern-Gerlach force on the motion of an electron in a plasma wave}

For clarity we use the simplified notation $t = C^0$, $x = C^1$, $y = C^2$, $z = C^3$ for the components of the worldline $C$.  Consider a system with the electromagnetic field 2-form $F$ associated with an electrostatic wave
%
%
%
\begin{align}
F_{ab} = \left\{\begin{array}{c}
E(\xi)\text{ for }a=0, b=3\\
-E(\xi)\text{ for }a=3, b=0\\
0\text{ otherwise},\\
\end{array}\right.\label{generalzwaveE}
\end{align}
where $\xi = z-vt$ is the phase of the wave.  The evolution equations \eqref{eqmotlinS1} are simply
\begin{align}
&\ddot{t} =  \left(1 + \frac{q_{\text{e}}}{m_{\text{e}}^2}S_{03}E\right)\frac{q_{\text{e}}}{m_{\text{e}}}E{\dot{z}}+\frac{q_{\text{e}}}{m_{\text{e}}^2}\left(v+\dot{t}\left(\dot{z}-v\dot{t}\right)\right)E'S_{03}\label{SGcomp1},\\
&\ddot{x} =  \frac{q_{\text{e}}}{m_{\text{e}}^2}\dot{x}\left(\dot{z}-v\dot{t}\right)E'S_{03}\label{SGcomp2},\\
&\ddot{y} =  \frac{q_{\text{e}}}{m_{\text{e}}^2}\dot{y}\left(\dot{z}-v\dot{t}\right)E'S_{03}\label{SGcomp3},\\
&\ddot{z} =  \left(1 + \frac{q_{\text{e}}}{m_{\text{e}}^2}S_{03}E\right)\frac{q_{\text{e}}}{m_{\text{e}}}E\dot{t}+\frac{q_{\text{e}}}{m_{\text{e}}^2}\left(1+\dot{z}\left(\dot{z}-v\dot{t}\right)\right)E'S_{03}\label{SGcomp4},
\end{align}
since $\frac{d}{d\tau}\dot{C}=\ddot{C}$ and $F_{03}=-F^{03}$. Note that dots denote derivatives with respect to the proper time $\tau$ and primes denote derivatives with respect to the phase $\xi$.  Similarly from \eqref{eqmotlinS2} the spin evolution equations are
\begin{align}
&\dot{S}_{01}=  \frac{q_{\text{e}}}{m_{\text{e}}}ES_{13},\quad\quad\dot{S}_{13}=  \frac{q_{\text{e}}}{m_{\text{e}}}ES_{01},\\
&\dot{S}_{02}=  \frac{q_{\text{e}}}{m_{\text{e}}}ES_{23},\quad\quad\dot{S}_{23}=  \frac{q_{\text{e}}}{m_{\text{e}}}ES_{02},\\
&\dot{S}_{03}=  0,~~~~\quad\quad\quad\quad\dot{S}_{12}=  0.\label{spincompconst}
\end{align}
Notably, the only component of the spin in \eqref{SGcomp1}-\eqref{SGcomp4} i.e. the only component that affects the trajectory of the particle is $S_{03}$, which according to \eqref{spincompconst} is constant. We hence neglect the remaining spin equations of motion when solving for the worldline of the electron. Writing the remaining equations of the system, \eqref{SGcomp1}-\eqref{SGcomp4}, in the coordinate system $\{\gamma\zeta,x,y,\gamma\xi\}$, adapted to an observer travelling with the plasma wave at speed $v$, where $\gamma = \frac{1}{\sqrt{1-v^2}}$ is the Lorentz factor of the wave and $\zeta = -t+vz$, we find
\begin{align}
\ddot{\zeta} &= \frac{q_{\text{e}}}{m_{\text{e}}^2}S_{03}E'\dot{\zeta}\dot{\xi} -\left(1 + \frac{q_{\text{e}}}{m_{\text{e}}^2}S_{03}E\right)\frac{q_{\text{e}}}{m_{\text{e}}}E{\dot{\xi}}\label{PWSPINEQOFMOTION1},\\
\ddot{\xi} &=\frac{q_{\text{e}}}{m_{\text{e}}^2}S_{03}E'\dot{\xi}\dot{\xi} -\left(1 + \frac{q_{\text{e}}}{m_{\text{e}}^2}S_{03}E\right)\frac{q_{\text{e}}}{m_{\text{e}}}E\dot{\zeta} + \frac{q_{\text{e}}}{m_{\text{e}}^2\gamma^2}S_{03}E',\label{PWSPINEQOFMOTION4}\\
\ddot{x} &=  \frac{q_{\text{e}}}{m_{\text{e}}^2}S_{03}E'\dot{x}\dot{\xi}\label{PWSPINEQOFMOTION2},\\
\ddot{y} &=  \frac{q_{\text{e}}}{m_{\text{e}}^2}S_{03}E'\dot{y}\dot{\xi}\label{PWSPINEQOFMOTION3}.
\end{align}
A particular solution to \eqref{PWSPINEQOFMOTION1}-\eqref{PWSPINEQOFMOTION3} includes a constant value for phase $\xi$ and has the form
\begin{align}
\zeta_{\text{sol}}(\tau)&=\frac{1}{\left(1+\frac{q_{\text{e}}}{m_{\text{e}}^2}S_{03}E_{\text{C}}\right)}\frac{S_{03}}{m_{\text{e}}\gamma^2}\frac{E'_{\text{C}}}{E_{\text{C}}}\tau + \zeta_0,\label{trajectoryzeta}\\
x_{\text{sol}}(\tau) &= \dot{x}_0\tau + x_0,\label{trajectoryx}\\
y_{\text{sol}}(\tau) &= \dot{y}_0\tau + y_0,\label{trajectoryy}\\
\xi_{\text{sol}} &= \xi_{\text{C}}\label{trajectotyxi},
\end{align}
and where $E_{\text{C}}$ denotes the value of the electric field at $\xi=\xi_{\text{C}}$. Here $x_0$, $y_0$, $\zeta_0$ are arbitrary constants and $\dot{x}_0$, $\dot{y}_0$ are arbitrary constants up to fulfilment of the  normalisation condition $g_{ab}\dot{C}^a\dot{C}^b = -1$ on the worldline of the electron, i.e.
\begin{align}
-\frac{1}{\left(1+\frac{q_{\text{e}}}{m_{\text{e}}^2}S_{03}E_{\text{C}}\right)^2}\left(\frac{S_{03}}{m_{\text{e}}\gamma^2}\frac{E'_{\text{C}}}{E_{\text{C}}}\right)^2 + \dot{x}^2_0 + \dot{y}^2_0=-1.\label{normalisationtrajectoryspin}
\end{align}
Note that condition \eqref{normalisationtrajectoryspin} places restrictions on the system parameters, for instance $S_{03}, E_{\text{C}}, E'_{\text{C}} \neq 0$. Consequently,  the solution family  \eqref{trajectoryzeta}-\eqref{trajectotyxi} does not exist for a spinless particle, nor a constant electromagnetic background. 
Also note that since we wish to consider a system with a large field gradient, we consider $\frac{E'_{\text{C}}}{E_{\text{C}}}$ to be of order $(S_{03})^{-1}$.
%
%

A sufficiently short and intense laser pulse propagating through a plasma may create a travelling longitudinal
plasma wave whose velocity is approximately the same as the laser pulse's group velocity. The electric field produced by such a plasma wave provides an excellent example of an electric field of the form \eqref{generalzwaveE}: 
\begin{align}
E = \frac{m_{\text{e}}\nu'}{q_{\text{e}}\gamma^2},\label{PWEintermsofnudef}
\end{align}
where $\nu$ must satisfy
\begin{align}
\frac{m_{\text{e}}^2}{2q_{\text{e}}^2\gamma^4}\nu'^2 - m_{\text{e}}Zn_{\text{ion}}\left(v\sqrt{\nu^2-\gamma^2}-\nu+\gamma\right) = 0\label{smallernucondition}
 \end{align}
in order to satisfy the Maxwell equations and the Lorentz force equation (see Appendix \ref{App:PW} for details).
Here $Z$ is the degree of ionisation, $n_{\text{ion}}$ is the proper number density of the (background) ions.
In laser wakefield acceleration, the ``target" is the dephasing point, where accelerated electrons begin to overtake the plasma wave. At this point the field gradient is much larger than the electric field, hence $\frac{E'_{\text{C}}}{E_{\text{C}}}$ can be said to be large (of order $(S_{03})^{-1}$). 

\begin{figure}
\centering
\includegraphics[scale=.2]{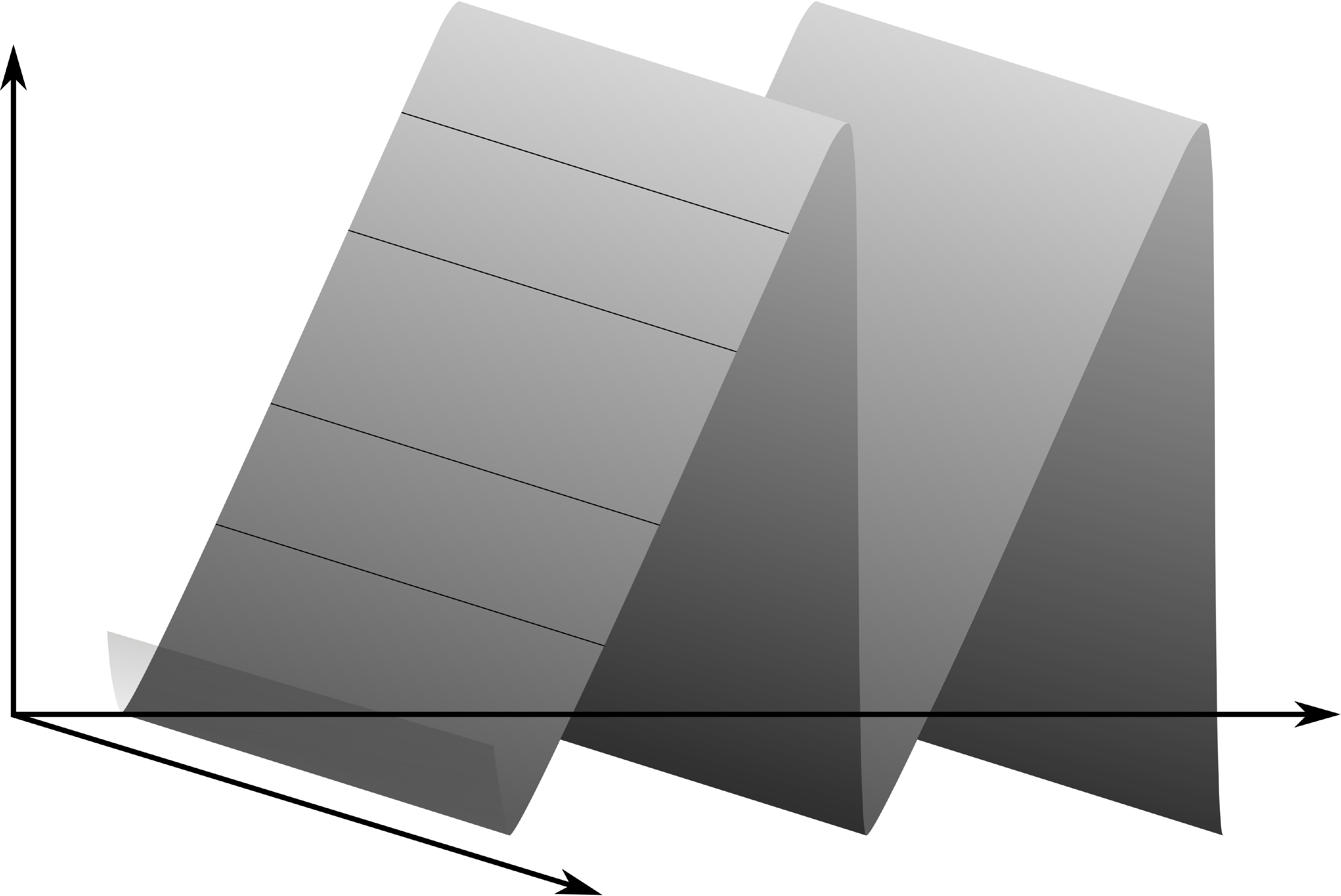}
\begin{picture}(0,0)
\put(-168,46){\scriptsize{$C_1$}}
\put(-159,65){\scriptsize{$C_2$}}
\put(-148,91){\scriptsize{$C_3$}}
\put(-140,108){\scriptsize{$C_4$}}
\put(-200,120){\small{$E$}}
\put(-10,15){\small{$\xi$}}
\put(-130,-5){\small{$x$}}
\end{picture}
\caption{Illustration of several example trajectories $C_1$, $C_2$, $C_3$, $C_4$  given by different choices of $\xi_{\text{C}}$. Whilst the plasma electrons travel along $\xi$, test electrons described by \eqref{trajectoryzeta}-\eqref{trajectotyxi}  travel transversely to the wave's velocity, `surfing' along the wave.}
\label{fig:PerturbationTrajectories}
\end{figure}
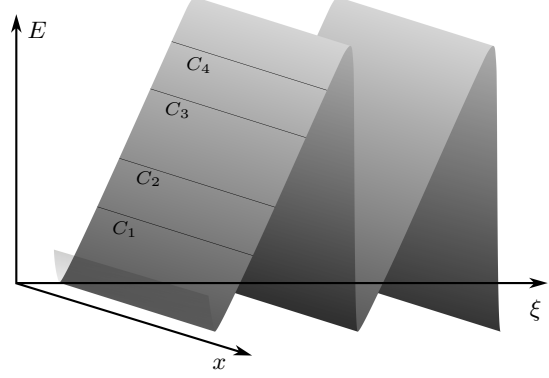

 The family of trajectories given by \eqref{trajectoryzeta}-\eqref{trajectotyxi} with the electric field \eqref{PWEintermsofnudef} is illustrated in Figure \ref{fig:PerturbationTrajectories}. Despite the propagation of the plasma electrons in the $\xi$ direction, the electrons described by the solution family \eqref{trajectoryzeta}-\eqref{trajectotyxi} travel transversely, along lines of constant $E$ (they `surf' along the wave).

\subsection{Stability of the `surfing' solutions in a plasma wave}
\label{stabilitysection}

It is natural now to consider the linear stability of the family of `surfing' solutions described by \eqref{trajectoryzeta}-\eqref{trajectotyxi} for the plasma wave electric field \eqref{PWEintermsofnudef}. In order to investigate this, consider the following:
\begin{align}
\zeta(\tau) &= \zeta_{\text{sol}}(\tau)+\varepsilon\Delta\zeta(\tau) \label{perturbedsol1},\\
x(\tau) &= x_{\text{sol}}(\tau) +\varepsilon\Delta x(\tau)\label{perturbedsol2},\\
y(\tau) &= y_{\text{sol}}(\tau) +\varepsilon\Delta y(\tau)\label{perturbedsol3},\\
\xi(\tau) &= \xi_{\text{sol}} +\varepsilon\Delta\xi(\tau) \label{perturbedsol4},
\end{align}
where $\varepsilon$ is a small constant and the $\Delta$ terms correspond to perturbations. Substituting \eqref{perturbedsol4} into $\nu$ and taking Taylor series in $\varepsilon$ gives:
\begin{align}
\nu(\xi_{\text{C}}+\varepsilon\Delta\xi) = \nu_C+\varepsilon\nu'_{\text{C}}\Delta\xi + \mathcal{O}(\varepsilon^2)\label{nuperturb},
\end{align}
where $\nu_C = \nu(\xi_C)$, $\left.\nu'_C= \frac{d\nu(\xi)}{d\xi}\right|_{\xi=\xi_C}$ and so on for the higher derivatives. Substituting \eqref{nuperturb}, its derivatives and \eqref{perturbedsol1}-\eqref{perturbedsol4} into the equations of motion \eqref{PWSPINEQOFMOTION1}-\eqref{PWSPINEQOFMOTION3} gives, to first order in the perturbations, 
\begin{align}
&\ddot{\Delta\zeta} =  \left[-\left(1 - \mathcal{S}\nu'_{\text{C}}\right)\frac{\nu'_{\text{C}}}{\gamma^2}+\frac{\mathcal{S}^2}{1-\mathcal{S}\nu'_{\text{C}}}\frac{(\nu''_{\text{C}})^2}{\nu'_{\text{C}}}\right]\dot{\Delta\xi}=  \mathcal{A}_1\dot{\Delta\xi},\label{PWSPINEQOFMOTION1firstorder}\\
&\ddot{\Delta x} =\left[-\mathcal{S}\dot{x}_0\nu''_{\text{C}}\right]\dot{\Delta\xi}=\mathcal{A}_2\dot{\Delta\xi},\label{PWSPINEQOFMOTION2firstorder}\\
&\ddot{\Delta y} =\left[-\mathcal{S}\dot{y}_0\nu''_{\text{C}}\right]\dot{\Delta\xi}=\mathcal{A}_3\dot{\Delta\xi},\label{PWSPINEQOFMOTION3firstorder}\\
&\ddot{\Delta \xi} = \left[(1-\mathcal{S}\nu'_{\text{C}})^2\frac{(\nu'_{\text{C}})^2}{\gamma^4}+\left(\frac{1-\mathcal{S}\nu'_{\text{C}}}{\nu'_{\text{C}}}-\frac{\mathcal{S}}{1-\mathcal{S}\nu'_{\text{C}}}\right)\frac{\mathcal{S}(\nu''_{\text{C}})^2}{\gamma^2}-\mathcal{S}\frac{\nu'''_{\text{C}}}{\gamma^2}\right]\Delta\xi=\mathcal{A}_4\Delta\xi,\label{PWSPINEQOFMOTION4firstorder}
\end{align}
where $\mathcal{S} = -\frac{S_{03}}{m_{\text{e}}\gamma^2}$ and the constants $\mathcal{A}_n$ depend on the spin, electric field and the plasma wave speed. A solution to \eqref{PWSPINEQOFMOTION1firstorder}-\eqref{PWSPINEQOFMOTION4firstorder} is 
\begin{align}
\Delta\zeta &= \frac{\mathcal{A}_1}{\sqrt{\mathcal{A}_4}}\left({\mathcal{C}_1}e^{\sqrt{\mathcal{A}_4}\tau} + {\mathcal{C}_2}e^{-\sqrt{\mathcal{A}_4}\tau}\right)\label{Deltaperturbedsol1},\\
\Delta x &= \frac{\mathcal{A}_2}{\sqrt{\mathcal{A}_4}}\left({\mathcal{C}_1}e^{\sqrt{\mathcal{A}_4}\tau} + {\mathcal{C}_2}e^{-\sqrt{\mathcal{A}_4}\tau}\right)\label{Deltaperturbedsol2},\\
\Delta y &= \frac{\mathcal{A}_3}{\sqrt{\mathcal{A}_4}}\left({\mathcal{C}_1}e^{\sqrt{\mathcal{A}_4}\tau} + {\mathcal{C}_2}e^{-\sqrt{\mathcal{A}_4}\tau}\right)\label{Deltaperturbedsol3},\\
\Delta\xi &=\mathcal{C}_1e^{\sqrt{\mathcal{A}_4}\tau} + \mathcal{C}_2e^{-\sqrt{\mathcal{A}_4}\tau},\label{Deltaperturbedsol4}
\end{align}
where $\mathcal{C}_1$, $\mathcal{C}_2$ are integration constants.  The stability of the system hence depends solely on the sign of the quantity $\mathcal{A}_4$, defined in \eqref{PWSPINEQOFMOTION4firstorder}. Written as a Taylor series in $\mathcal{S}$, $\mathcal{A}_4$ can be expressed as 
\begin{align}
{\mathcal{A}}_4 &= \left(\frac{\nu'_{\text{C}}}{\gamma^2}\right)^2 + \left(\left(\frac{\nu''_{\text{C}}}{\nu'_{\text{C}}}\right)^2-2\left(\frac{\nu'_{\text{C}}}{\gamma}\right)^2-\frac{\nu'''_{\text{C}}}{\nu'_{\text{C}}}\right)\frac{\mathcal{S}\nu'_{\text{C}}}{\gamma^2} + \mathcal{O}(\mathcal{S}^2)\label{linearisedA4}.
\end{align}
Assuming that the zeroth and first order terms are dominant we neglect the higher order terms\footnote{The range of validity of this assumption is ascertained in the subsequent section.}. Consequently, the exponential terms in the perturbations \eqref{Deltaperturbedsol1}-\eqref{Deltaperturbedsol4} become
\begin{align}
e^{\sqrt{\mathcal{A}_4}}\tau \approx&  \left(1+N_{\text{C}}\mathcal{S}\tau\right)e^{\frac{\left|\nu'_{\text{C}}\right|}{\gamma^2}\tau},\label{expassumption}
\end{align}
\label{sym:NC}to first order in $\mathcal{S}$, where
\begin{align}
N_{\text{C}} = \frac{1}{2}\frac{\nu'_{\text{C}}}{|\nu'_{\text{C}}|}\left(\left(\frac{\nu''_{\text{C}}}{\nu'_{\text{C}}}\right)^2-2\left(\frac{\nu'_{\text{C}}}{\gamma}\right)^2-\frac{\nu'''_{\text{C}}}{\nu'_{\text{C}}}\right).
\end{align}
Hence the $\xi$ perturbation \eqref{perturbedsol4} to first order in $\mathcal{S}$ is
\begin{align}
\Delta \xi = \mathcal{C}_1\left(1+N_{\text{C}}\mathcal{S}\tau\right)e^{\frac{\left|q_{\text{e}}E_{\text{C}}\right|}{m_{\text{e}}}\tau} + \mathcal{C}_2\left(1-N_{\text{C}}\mathcal{S}\tau\right)e^{-\frac{\left|q_{\text{e}}E_{\text{C}}\right|}{m_{\text{e}}}\tau},
\end{align}
that is the perturbation $\Delta\xi$ is unstable (to first order in $\mathcal{S}$) as the first exponential will diverge as $\tau$ increases, unless $\mathcal{C}_1=0$. Since the other three perturbations are closely linked to $\Delta\xi$, the complete perturbation is also divergent (unless the integration constant $\mathcal{C}_1=0$).

\subsection{The range of validity of {\eqref{expassumption}}}

In order to confirm that the zeroth and first order terms of $\mathcal{S}$ in $\mathcal{A}_4$ are dominant, consider the ratio $R$ of the sum of the zeroth and first order terms to the full expression for $\mathcal{A}_4$ \eqref{PWSPINEQOFMOTION4firstorder}. The ratio $R$ can be written as
\begin{align}
R = \frac{\hat{\nu'}}{(1-\hat{\nu'})}\left[\frac{(\hat{\nu'})^4-2\gamma^2(\hat{\nu''})^2 +\hat{\nu'}\left(\gamma^2(\hat{\nu''})^2-(\hat{\nu'})^4\right)}{(\hat{\nu'})^3 + \gamma^2(\hat{\nu''})^2-2(\hat{\nu'})^4-\gamma^2\hat{\nu'''}\hat{\nu'}}\right],\label{RelativeR2}
\end{align}
where $\hat{\nu'} = \frac{d}{d({\mathcal{S}^{-1}\xi})}\nu(\xi)$.  

Returning to SI units for clarity, introducing the Schwinger limit $E_{\text{S}} = \frac{m_{\text{e}}^2c^3}{q_{\text{e}}\hbar}$\label{sym:Eschwinger} and the maximum electric field (the wave breaking limit) for a cold plasma, $E_{\text{\text{max}}} = c\sqrt{\frac{2(\gamma-1)m_{\text{e}}Zn_{\text{ion}}}{\epsilon_0}}$\label{sym:EmaxPW}, allows the convenient re-parameterisation of the system in terms of the free parameters $\{v, \hat{\nu'}, k\}$ where
\begin{align}
\hat{\nu'} &= -\frac{E}{E_{\text{S}}}\frac{S_{03}}{\hbar},\\
k  &= \left(\frac{E_{\text{\text{max}}}}{E_{\text{S}}}\right)^2\left(\frac{S_{03}}{\hbar}\right)^2.
\end{align}
%
The Schwinger limit characterises the electric field strength at which quantum vacuum effects are expected to be significant \cite{Piazza}
. The present analysis does not take such phenomena into account and therefore the conditions $E<E_{\rm S}$, $E_{\rm max}<E_{\rm S}$ are required. Thus, in addition to $0<v<1$, it follows that $-1<\hat{\nu'}<1$ and $0<k<1$. We also have the restriction
\begin{align}
\frac{(\hat{\nu'})^2}{k} = \left(\frac{E}{E_{\text{max}}}\right)^2 <  1\label{paramspacelimit},
\end{align}
to ensure that $E < E_{\text{max}}$.
The combination of these conditions also guarantees that $\nu > \gamma$, ensuring that the square root in \eqref{smallernucondition} remains real. The maximum amplitude plasma equation \eqref{smallernucondition} and its derivatives relate $\hat{\nu'}$ and its derivatives:
\begin{align}
\hat{\nu}_\pm &= -\gamma^2\left((\gamma-1)\frac{(\hat{\nu'})^2}{k}-\gamma\right)\pm\gamma^2v\sqrt{\left((\gamma-1)\frac{(\hat{\nu'})^2}{k}-\gamma\right)^2-1}\label{nunondim2},\\
\hat{\nu''} &= \frac{1}{2}\frac{k}{\gamma-1}\left(v\frac{\hat{\nu}_\pm}{\sqrt{\hat{\nu}_\pm^2-\gamma^2}}-1\right)\label{dnunondim2},\\
\hat{\nu'''} &= -\frac{1}{2}\frac{k}{\gamma-1}v\gamma^2\frac{\hat{\nu'}}{(\hat{\nu}_\pm^2-\gamma^2)^{3/2}}\label{d2nunondim2}.
\end{align}

\begin{figure*}
\centering
\includegraphics[scale=.3]{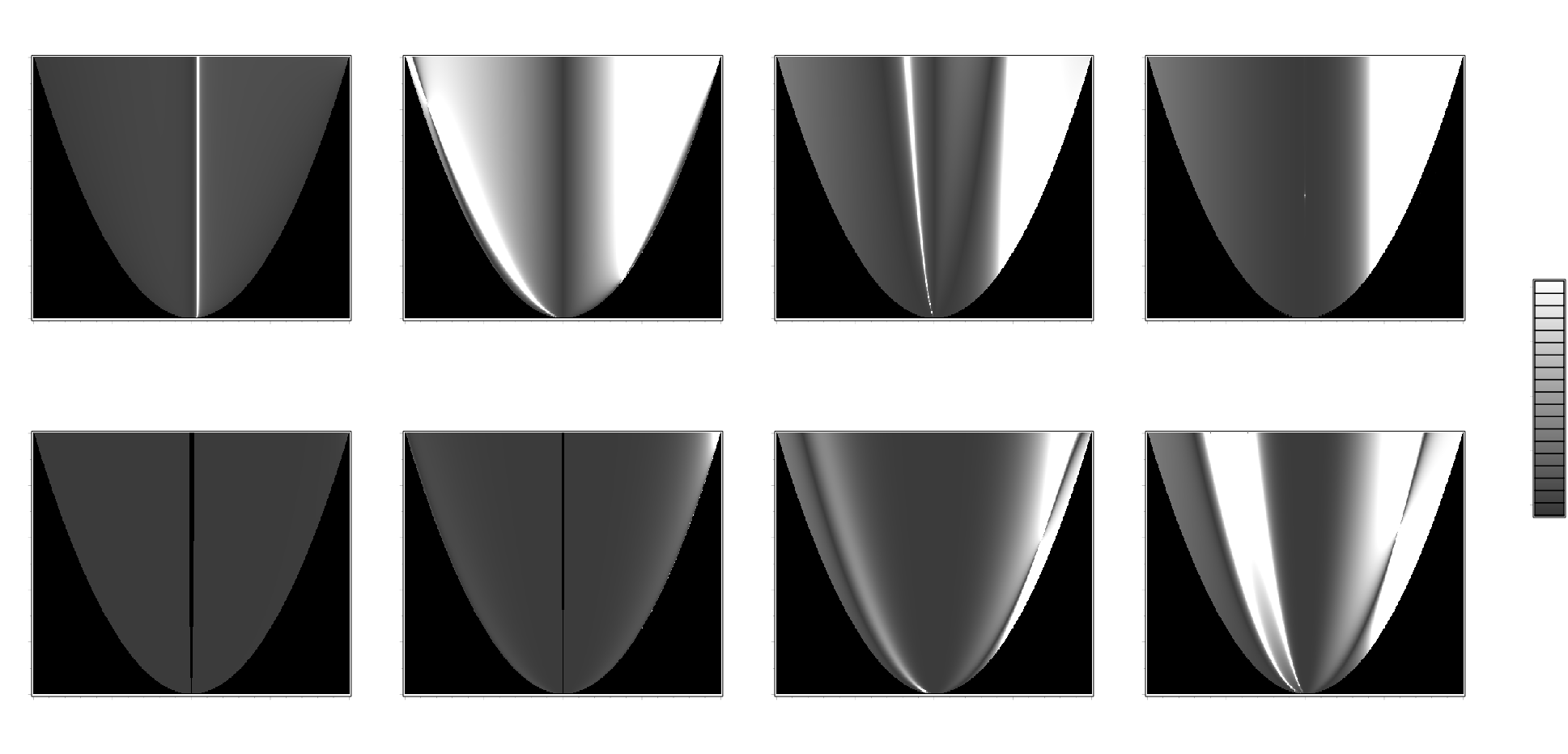}
\begin{picture}(0,0)
%
%
%
%
%
\put(-435,8){\scriptsize{-1}}
\put(-389,8){\scriptsize{0}}
\put(-346,8){\scriptsize{1}}
\put(-331,8){\scriptsize{-1}}
\put(-285,8){\scriptsize{0}}
\put(-242,8){\scriptsize{1}}
\put(-227,8){\scriptsize{-1}}
\put(-181,8){\scriptsize{0}}
\put(-138,8){\scriptsize{1}}
\put(-123,8){\scriptsize{-1}}
\put(-77,8){\scriptsize{0}}
\put(-34,8){\scriptsize{1}}
\put(-435,113){\scriptsize{-1}}
\put(-389,113){\scriptsize{0}}
\put(-346,113){\scriptsize{1}}
\put(-331,113){\scriptsize{-1}}
\put(-285,113){\scriptsize{0}}
\put(-242,113){\scriptsize{1}}
\put(-227,113){\scriptsize{-1}}
\put(-181,113){\scriptsize{0}}
\put(-138,113){\scriptsize{1}}
\put(-123,113){\scriptsize{-1}}
\put(-77,113){\scriptsize{0}}
\put(-34,113){\scriptsize{1}}
%
%
%
\put(-438,14){\scriptsize{0}}
\put(-438,86){\scriptsize{1}}
\put(-334,14){\scriptsize{0}}
\put(-334,86){\scriptsize{1}}
\put(-230,14){\scriptsize{0}}
\put(-230,86){\scriptsize{1}}
\put(-126,14){\scriptsize{0}}
\put(-126,86){\scriptsize{1}}
\put(-438,119){\scriptsize{0}}
\put(-438,191){\scriptsize{1}}
\put(-334,119){\scriptsize{0}}
\put(-334,191){\scriptsize{1}}
\put(-230,119){\scriptsize{0}}
\put(-230,191){\scriptsize{1}}
\put(-126,119){\scriptsize{0}}
\put(-126,191){\scriptsize{1}}
%
%
%
\put(-389,-2){\scriptsize{$\hat{\nu'}$}}
\put(-285,-2){\scriptsize{$\hat{\nu'}$}}
\put(-181,-2){\scriptsize{$\hat{\nu'}$}}
\put(-77,-2){\scriptsize{$\hat{\nu'}$}}
\put(-389,103){\scriptsize{$\hat{\nu'}$}}
\put(-285,103){\scriptsize{$\hat{\nu'}$}}
\put(-181,103){\scriptsize{$\hat{\nu'}$}}
\put(-77,103){\scriptsize{$\hat{\nu'}$}}
%
%
%
\put(-440,51){\scriptsize{$k$}}
\put(-336,51){\scriptsize{$k$}}
\put(-232,51){\scriptsize{$k$}}
\put(-128,51){\scriptsize{$k$}}
\put(-440,156){\scriptsize{$k$}}
\put(-336,156){\scriptsize{$k$}}
\put(-232,156){\scriptsize{$k$}}
\put(-128,156){\scriptsize{$k$}}
%
%
%
\put(0,98){\scriptsize{$|R|$}}
\put(-3,128){\scriptsize{$1$}}
\put(-3,63){\scriptsize{$0$}}
%
%
%
\put(-399,203){\scriptsize{$v$ = $0.1c$}}
\put(-295,203){\scriptsize{$v$ = $0.5c$}}
\put(-191,203){\scriptsize{$v$ = $0.9c$}}
\put(-87,203){\scriptsize{$v$ = $0.999c$}}
%
%
%
\put(-338,208){\line(0,-1){210}}
\put(-234,208){\line(0,-1){210}}
\put(-130,208){\line(0,-1){210}}
\end{picture}
\caption{The relative size of terms $R$ in the parameter space ($\hat{\nu'}$, $k$) for four values of plasma wave speed $v$. For each speed there are two plots, the upper using $\hat{\nu}$ \eqref{nunondim2} with the positive sign and the lower using the negative sign. The regions in which the trajectories of `surfing' particles \eqref{trajectoryzeta}-\eqref{trajectotyxi} are linearly unstable (i.e. the assumptions made in Section \ref{stabilitysection} are valid) are the regions where $|R| \ll 1$.}
\label{fig:parameterspaceplots}
\end{figure*}
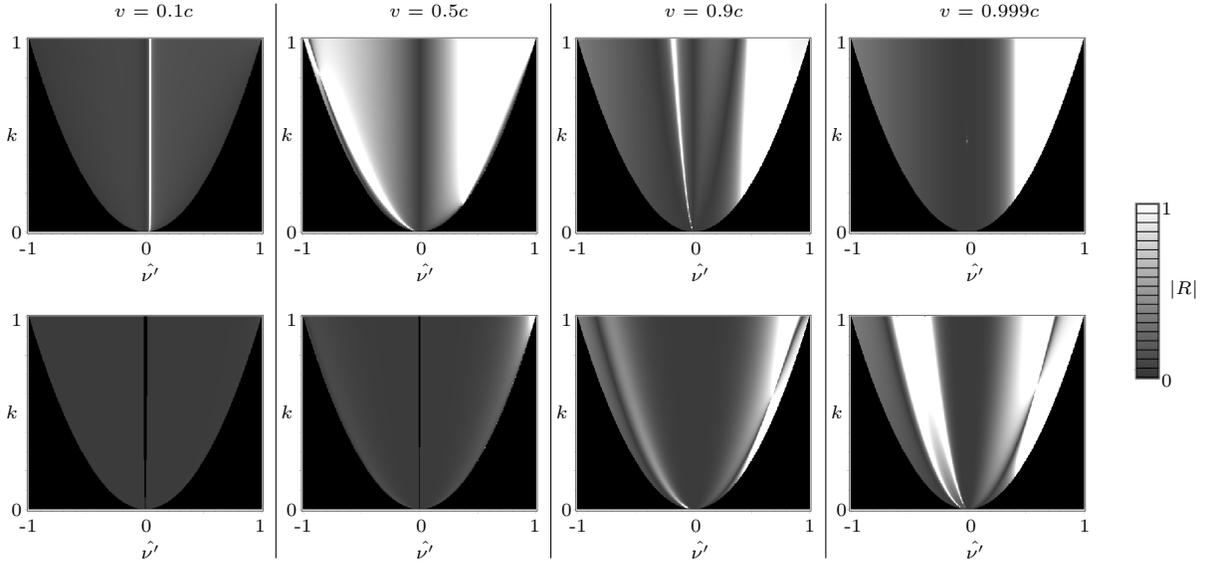

Figure \ref{fig:parameterspaceplots} illustrates the size of \eqref{RelativeR2} across the parameter space ($\hat{\nu'}$, $k$) --  it is clear that $|R|\ll 1$ for the majority of the parameter space. Several things should be made clear, however.  Firstly, the black region in each plot is excluded by the condition \eqref{paramspacelimit} -- these dark regions correspond to electric fields $E > E_{\text{max}}$. The central line present in some plots indicates that the electric field must be non-zero, as is already stipulated by the normalisation condition \eqref{normalisationtrajectoryspin}. Secondly, the result of the numerical analysis is unreliable in certain regions of the plots, for example near the edge of the parabola or along the line $\hat{\nu'} = 0$.  

\begin{figure*}
\centering
\includegraphics[scale=.345]{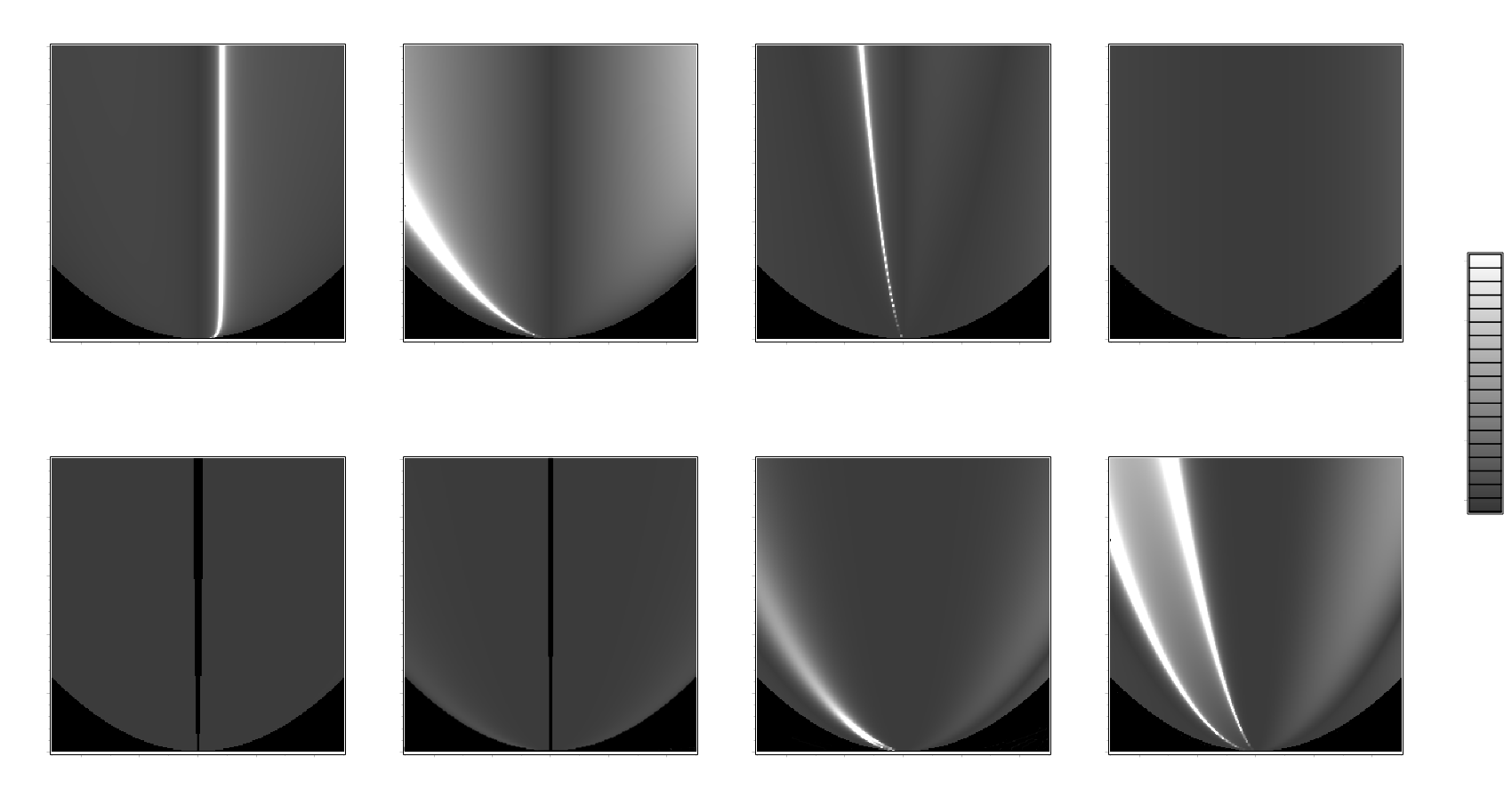}
\begin{picture}(0,0)
%
%
%
%
%
\put(-435,5){\scriptsize{-0.25}}
\put(-388,5){\scriptsize{0}}
\put(-355,5){\scriptsize{0.25}}
\put(-331,5){\scriptsize{-0.25}}
\put(-285,5){\scriptsize{0}}
\put(-251,5){\scriptsize{0.25}}
\put(-227,5){\scriptsize{-0.25}}
\put(-182,5){\scriptsize{0}}
\put(-148,5){\scriptsize{0.25}}
\put(-123,5){\scriptsize{-0.25}}
\put(-79,5){\scriptsize{0}}
\put(-45,5){\scriptsize{0.25}}
\put(-435,126){\scriptsize{-0.25}}
\put(-388,126){\scriptsize{0}}
\put(-355,126){\scriptsize{0.25}}
\put(-331,126){\scriptsize{-0.25}}
\put(-285,126){\scriptsize{0}}
\put(-251,126){\scriptsize{0.25}}
\put(-227,126){\scriptsize{-0.25}}
\put(-182,126){\scriptsize{0}}
\put(-148,126){\scriptsize{0.25}}
\put(-123,126){\scriptsize{-0.25}}
\put(-79,126){\scriptsize{0}}
\put(-45,126){\scriptsize{0.25}}
%
%
%
\put(-435,12){\scriptsize{0}}
\put(-435,100){\scriptsize{0.25}}
\put(-334,12){\scriptsize{0}}
\put(-334,100){\scriptsize{0.25}}
\put(-230,12){\scriptsize{0}}
\put(-230,100){\scriptsize{0.25}}
\put(-126,12){\scriptsize{0}}
\put(-126,100){\scriptsize{0.25}}
\put(-435,133){\scriptsize{0}}
\put(-435,221){\scriptsize{0.25}}
\put(-334,133){\scriptsize{0}}
\put(-334,221){\scriptsize{0.25}}
\put(-230,133){\scriptsize{0}}
\put(-230,221){\scriptsize{0.25}}
\put(-126,133){\scriptsize{0}}
\put(-126,221){\scriptsize{0.25}}
%
%
%
\put(-388,-5){\scriptsize{$\hat{\nu'}$}}
\put(-285,-5){\scriptsize{$\hat{\nu'}$}}
\put(-182,-5){\scriptsize{$\hat{\nu'}$}}
\put(-79,-5){\scriptsize{$\hat{\nu'}$}}
\put(-388,115){\scriptsize{$\hat{\nu'}$}}
\put(-285,115){\scriptsize{$\hat{\nu'}$}}
\put(-182,115){\scriptsize{$\hat{\nu'}$}}
\put(-79,115){\scriptsize{$\hat{\nu'}$}}
%
%
%
\put(-440,55){\scriptsize{$k$}}
\put(-336,55){\scriptsize{$k$}}
\put(-232,55){\scriptsize{$k$}}
\put(-128,55){\scriptsize{$k$}}
\put(-440,175){\scriptsize{$k$}}
\put(-336,175){\scriptsize{$k$}}
\put(-232,175){\scriptsize{$k$}}
\put(-128,175){\scriptsize{$k$}}
%
%
%
\put(0,120){\scriptsize{$|R|$}}
\put(-3,155){\scriptsize{$1$}}
\put(-3,80){\scriptsize{$0$}}
%
%
%
\put(-399,230){\scriptsize{$v$ = $0.1c$}}
\put(-295,230){\scriptsize{$v$ = $0.5c$}}
\put(-191,230){\scriptsize{$v$ = $0.9c$}}
\put(-87,230){\scriptsize{$v$ = $0.999c$}}
%
%
%
\put(-338,235){\line(0,-1){240}}
\put(-235,235){\line(0,-1){240}}
\put(-132,235){\line(0,-1){240}}
\end{picture}
\caption{The relative size of terms $R$ in the parameter space ($\hat{\nu'}$, $k$) for four values of plasma wave speed $v$. For each speed there are two plots, the upper using $\hat{\nu}$ \eqref{nunondim2} with the positive sign and the lower using the negative sign. The regions in which the trajectories \eqref{trajectoryzeta}-\eqref{trajectotyxi} of `surfing' particles are linearly unstable (i.e. the assumptions made in Section \ref{stabilitysection} are valid) are the regions where $|R| \ll 1$.}
\label{fig:parameterspaceplotszoom}
\end{figure*}
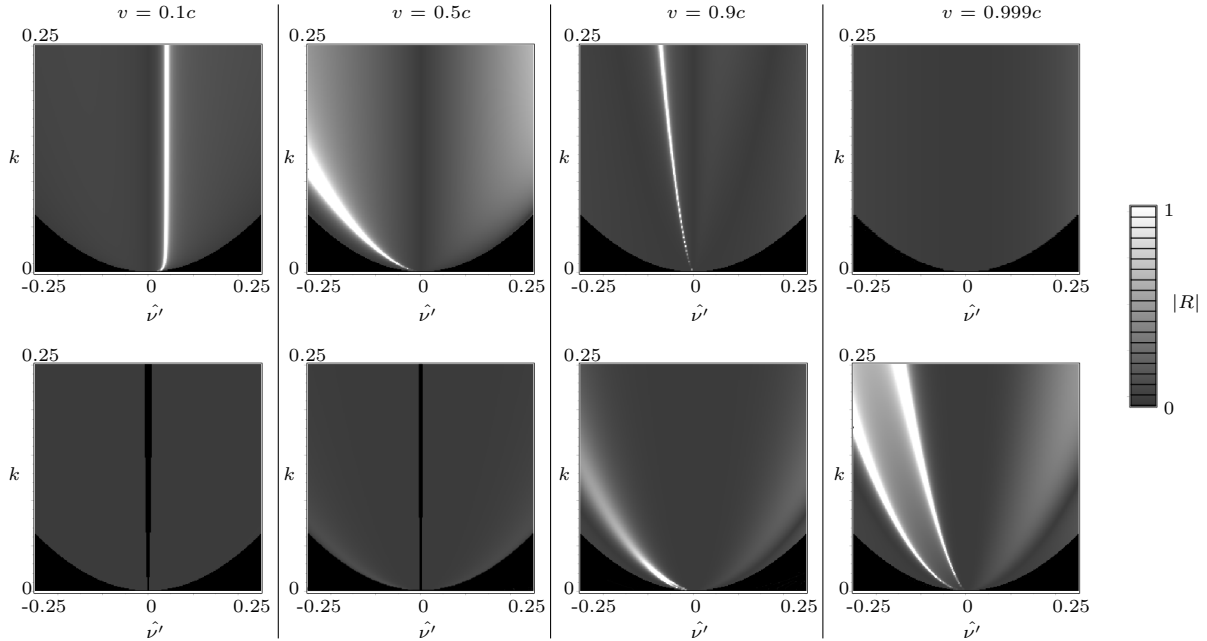

For values of the electric field $E$ of the plasma wave (and the maximum electric field of the cold plasma $E_{\text{max}}$) that are much less than the Schwinger limit $E_{\text{S}}$, both $|\hat{\nu'}|$ and $k$ are much less than 1 -- hence the crucial regions of the parameter space to consider are relatively close to the origin. Figure \ref{fig:parameterspaceplotszoom} illustrates the size of \eqref{RelativeR2} across a smaller range of electric fields. Here in almost every plot the relative size of terms \eqref{RelativeR2} is typically much less than 1, and hence the assumption \eqref{expassumption} is generally valid. Hence it is safe to conclude that the trajectories described by \eqref{trajectoryzeta}-\eqref{trajectotyxi} corresponding to the `surfing' particles are linearly unstable across the vast majority of the parameter space for $E \ll E_{\text{S}}$.

\section{Conclusion}

A new family of trajectories for a classical charged particle with spin in an electrostatic plasma wave has been presented -- notably these trajectories do not exist for a non-spinning particle, nor for a non-varying electric field. These trajectories represent particles moving transverse to the wave propagation, `surfing' along the wave. The linear stability of these trajectories depends on the values of the plasma wave speed, the electric field and the spin component $S_{03}$. As shown in Section \ref{stabilitysection}, these trajectories are linearly unstable for the majority of the parameter space, though there are some small regions in the parameter space where this may not be the case, where the assumptions of \eqref{expassumption} are invalid. For lower electric field, as in Figure \ref{fig:parameterspaceplotszoom}, these regions are even less prominent.

The existence of transverse trajectories has adverse consequences for the size of electron bunches in laser wakefield accelerators; electrons may slip into and out of these transverse trajectories once they catch up with the wave, spreading out into a disc oriented with normal parallel to the wave propagation direction. The linear instability of these solutions, however, implies that any electron that enters a transverse trajectory would likely leave it soon afterwards. 

It is important to note, however, that this is an instance in which the spin of a particle affects its trajectory more than radiation reaction effects -- indeed in this instance, since the particles in the new trajectories are travelling at constant speed, the radiation reaction effects are negligible. Hence we recommend that researchers aiming to model such systems consider including spin effects in their PIC codes.  

\section{Acknowledgements}

This work was undertaken as part of the ALPHA-X consortium funded under EPSRC grant EP/J018171/1 and with support from the Cockcroft Institute of Accelerator Science and Technology (STFC grant ST/G008248/1). S.P. Flood was funded by an EPSRC studentship. The authors thank Adam Noble for helpful comments on the manuscript.

\appendix
\section{Appendix: Derivation of the equations of motion}
\label{app:derivation}
\emph{This Appendix makes use of Euclidean $3$-vectors, $4$-vectors on Minkowski spacetime and de Rham currents (Schwarz distributions on differential forms). For clarity, a different notation will be used for each of those objects. Use of an arrow $\vec{V}$ indicates a Euclidean $3$-vector, $V$ is a $4$-vector with the appropriate metric dual $\widetilde{V}$, and a subscript D such as $V_D$ indicates a de Rham current.}

{~}

This Appendix shows a new derivation of the relativistic Stern-Gerlach force and TBMT equations for a charged particle, using exterior calculus and Schwarz distributions\footnote{For a summary of the conventions used in this Appendix, see Ref. \cite{BurtonPrimer}}. The aim of this section is to reduce the local balance laws \cite{burton2008spinning}
\begin{align}
d\mathcal{T}^a &= i_{X^a}F\wedge j^{\text{free}} + i_{X^a}F\wedge j^{\text{bound}},\label{uberbalancelaw}\\ 
d\sigma^{ab} &= \frac{1}{2}\left(dx^a\wedge\mathcal{T}^b - dx^b\wedge\mathcal{T}^a\right)\label{uberbalancelaw2},
\end{align}
for a charged continuum, with stress 3-forms $\mathcal{T}^a$ and spin 3-forms $\sigma^{ab}$, to a particle model using de Rham currents.  The vector basis $\{X^a\}$ is a Killing frame, and the electromagnetic 2-form $F$ and current 3-forms $j^{\text{free}}$, $j^{\text{bound}}$ satisfy the Maxwell equations
\begin{align}
dF &= 0,\\
d\star F &=  j^{\text{free}} + j^{\text{bound}}.\label{Max2}
\end{align}

Given a charged fluid, the magnetisation and polarisation vectors are given by
\begin{align}
\vec{\mathcal{P}}(\vec{r},t)=n(\vec{r},t)\vec{\mu}_{\text{e}}(\vec{r},t),\\
\vec{\mathcal{M}}(\vec{r},t)=n(\vec{r},t)\vec{\mu}_{\text{m}}(\vec{r},t),
\end{align}
where $n$ is the particle number density and $\vec{\mu}_{\text{e}}$ and $\vec{\mu}_{\text{m}}$ are the electric and magnetic dipole moments respectively. 

In order to efficiently move from the continuum model to a single-particle model, de Rham currents \cite{derham} are introduced. Firstly, in order to establish the notation in a simple setting, it is assumed that the fluid is at rest and hence described by a 4-vector field given by ${V}=\partial_t$. Then the distributional current associated with the worldline of a particle is introduced via 
\begin{align}
\int_\mathcal{M}\hat{f}n\star 1 \rightarrow \int_C\hat{f}dt,\label{fluidtoparticletrick}
\end{align}
analogous identifying the particle density as a Dirac delta function in order to reduce the domain of the integral to the particle's worldline $C$. The orientation is chosen so that the spacetime volume is $\star1 = dt\wedge dx\wedge dy\wedge dz$.  In \eqref{fluidtoparticletrick}, the curve $C$ has constant $x,y,z$ (due to the temporary choice of $V$) and $\hat{f}$ is a test function. Since the aim of this method is to induce the equation of motion of a particle from a fluid description, $C$ is assumed to be an integral curve of ${V}$. 

In order to find the appropriate distributions for the particle versions of the magnetisation and polarisation, consider the following. Given a Killing 3-vector $\vec{K} \in \{\vec{i},\vec{j},\vec{k}\}$, where $\vec{i}\cdot\vec{i}=1$, $\vec{j}\cdot\vec{j}=1$, $\vec{k}\cdot\vec{k}=1$ and $\vec{i}\cdot\vec{j}=\vec{i}\cdot\vec{k}=\vec{j}\cdot\vec{k}=0$\label{sym:3vecbasis}\label{sym:3vecK}, it is natural to introduce the de Rham current $\left(\vec{P}\cdot\vec{K}\right)_D$ as
\begin{align}
\left(\vec{\mathcal{P}}\cdot\vec{K}\right)_D[\hat{f}\star1] &= \int_C\vec{\mu}_{\text{e}}\cdot\vec{K}\hat{f}dt,\end{align}
where $\cdot$ represents the usual scalar product on 3-vectors and the subscript $\left(\vec{\mathcal{P}}\cdot\vec{K}\right)_D$ represents the distribution associated with the scalar $\vec{\mathcal{P}}\cdot\vec{K}$ (likewise for other quantities).  Expanding on the first of these equations:
\begin{align}
\left(\vec{\mathcal{P}}\cdot\vec{K}\right)_D\star 1[\hat{f}] &= \int_C\vec{\mu}_{\text{e}}\cdot\vec{K}\hat{f}dt\nonumber\\
&= C_D[\vec{\mu}_{\text{e}}\cdot\vec{K}\hat{f}dt]
\end{align}
and since $C_D[\alpha\hat{f}] = \left(C_D\wedge\alpha\right)[\hat{f}]$ for any 3-form $\alpha$. Since the form degree of $C_D$ is 3, $C_D\wedge dt = -dt\wedge C_D$ and hence 
\begin{align}
\left(\vec{\mathcal{P}}\cdot\vec{K}\right)_D\star 1[\hat{f}] &= -\left(\vec{\mu}_{\text{e}}\cdot\vec{K}\right)\left(dt\wedge C_D\right)[\hat{f}].
\end{align}
Stripping off the test function and noting that in this case $\dot{C}=\partial_t$ and $\star\star 1=-1$:
\begin{align}
\left(\vec{\mathcal{P}}\cdot\vec{K}\right)_D &= -\left(\vec{\mu}_{\text{e}}\cdot\vec{K}\right)\star\left(\widetilde{\dot{C}}\wedge C_D\right).\label{disteqold514}
\end{align}
Introducing the 4-vector $\mu_{\text{e}} = \mu_{{\text{e}}x}\partial_x + \mu_{{\text{e}}y}\partial_y + \mu_{{\text{e}}z}\partial_z$, where ${\mu}_{{\text{e}}x}$ is the $x$-component of the vector $\vec{\mu}_{\text{e}}$ etc., it follows that the polarisation of the particle given as $\mathcal{P}_D = \left(\vec{\mathcal{P}}\cdot\vec{K}_A\right)_Ddx^A$ is
\begin{align}
\mathcal{P}_D &=  \left(i_{\dot{C}}\star C_D\right)\widetilde{\mu_{\text{e}}}\label{poldist01},
\end{align}
for $A\in \{1,2,3\}$ and similarly the magnetisation $\mathcal{M}_D$ of the particle is
\begin{align}
\mathcal{M}_D &= \left(i_{\dot{C}}\star C_D\right)\widetilde{\mu_{\text{m}}}\label{poldist02}.
\end{align}
From now on, the worldline $C$ of the particle is taken to be arbitrary and $\mu_{\text{e}}$ and $\mu_{\text{m}}$ are 4-vectors orthogonal to the 4-velocity $\dot{C}$ of the particle. Furthermore, the timelike unit normalised 4-vector field $V$ is $\dot{C}$ when evaluated over the image of $C$.

The polarisation and magnetisation can be incorporated into a single 2-form, the polarisation 2-form $\Pi = G-F$, where $G$ is the excitation 2-form, given by:
\begin{align}
\Pi = -\widetilde{V}\wedge \widetilde{\mathcal{P}} + \star(\widetilde{V}\wedge\widetilde{\mathcal{M}}),\label{Polarisation2form}
\end{align}
where ${V}$ is the 4-vector describing the motion of the fluid. Using \eqref{poldist01} and \eqref{poldist02}, the distributional analogue of the polarisation 2-form can be written 
\begin{align}
\Pi_D &= \star C_D\wedge \widetilde{\mu_{\text{e}}} - \star(\star C_D\wedge\widetilde{\mu_{\text{m}}})\\
&=\star C_D\wedge \widetilde{\mu_{\text{e}}} - i_{\mu_{\text{m}}}C_D.
\end{align}

In order to establish a distributional analogue of the balance law \eqref{uberbalancelaw}, the distributional forms of the currents $j^{\text{free}}$ and $j^{\text{bound}}$ must be formulated. The free current is found by noting that
\begin{align}
j^{\text{free}} = -qn\star\widetilde{V},
\end{align}
where $q$ is the electron's charge, and therefore via \eqref{fluidtoparticletrick} we identify 
\begin{align}
j^{\text{free}}_D=qC_D. \label{jfreeqcd}
\end{align}
Since the excitation form $G$ satisfies $d\star G = j^{\text{free}}$, the field equation \eqref{Max2} shows that the bound current must satisfy $j^{\text{bound}} = -d\star \Pi$. Hence $j^{\text{bound}}_D = -d\star \Pi_D$ and the balance law \eqref{uberbalancelaw} can be written
\begin{align}
d\mathcal{T}^a_D = \int^{\tau_{\text{max}}}_{\tau_{\text{min}}}\left\{-qi_{\dot{C}}i_{X^a}F\hat{f} + \star\left(d\hat{f}\wedge i_{X^a}{F}\wedge\star\Sigma\right) + \star\left(\mathcal{L}_{X^a}{F}\wedge\star\Sigma\right)\hat{f}\right\}d\tau,
\end{align}
where $\{X^a\}$ is a basis of translational Killing vectors and the 2-form $\Sigma$ is defined as 
\begin{align}
\Sigma=-\widetilde{\dot{C}}\wedge\widetilde{\mu_{\text{e}}}+\star(\widetilde{\dot{C}}\wedge\widetilde{\mu_{\text{m}}}).
\end{align}
Further simplification occurs if we split $d\hat{f}$ into its $\dot{C}$-parallel and $\dot{C}$-orthogonal pieces, leaving 
\begin{align}
d\mathcal{T}^a_D[\hat{f}] &= \int^{\tau_{\text{max}}}_{\tau_{\text{min}}}\left\{\left[-qi_{\dot{C}}i_{X^a}{F}-\Sigma\cdot\left(\mathcal{L}_{X^a}{F}\right) - \nabla_{\dot{C}}((i_{X^a}{F}) \cdot i_{\dot{C}}\Sigma)\right]\hat{f}\right.\nonumber\\
&~~~~~\left.\quad\quad\quad\quad+ \star\left(\Pi^\perp_{\dot{C}}d\hat{f}\wedge i_{X^a}{F}\wedge\star\Sigma\right)\right\}d\tau.\label{bigdtauk}
\end{align}
We now choose the stress distribution ansatz in order to satisfy this expression. Allowing $\mathcal{T}^a_D$ to be of the form
\begin{align}
\mathcal{T}^a_D = -g(\mathbf{\pi},{X^a})C_D + i_{\dot{C}}\left(i_{X^a}{F}\wedge\star\Sigma\right)\wedge\star C_D\label{taukeq},
\end{align}
where $\pi$ is a 4-momentum vector, with the second term designed to absorb the worldline-orthogonal pieces of \eqref{bigdtauk}, the stress balance law can be written in the simple form
\begin{align}
\nabla_{\dot{C}}\left(i_{X^a}\widetilde{\pi}+ (i_{X^a}{F}) \cdot i_{\dot{C}}\Sigma\right)= -qi_{\dot{C}}i_{X^a}{F}-\Sigma\cdot\left(\mathcal{L}_{X^a}{F}\right),\label{Peqofmotion1}
\end{align}
where $\nabla$ is the Levi-Civita connection. Similarly, the spin balance equation \eqref{uberbalancelaw2}, upon substitution of \eqref{taukeq} becomes
\begin{align}
d\sigma^{ab}_D[\hat{f}] = \int^{\tau_{\text{max}}}_{\tau_{\text{min}}}\frac{1}{2}&\left[\dot{C}^a\left({\pi}^b+(i_{X^b}{F})\cdot i_{\dot{C}}\Sigma\right)+(i_{X^b}{F})\cdot i_{X^a}\Sigma\right.\nonumber\\
  &\quad-\dot{C}^b\left({\pi}^a+(i_{X^a}{F})\cdot i_{\dot{C}}\Sigma\right)-(i_{X^a}{F})\cdot i_{X^b}\Sigma\Big]\hat{f}d\tau,
\end{align}
and choosing the ansatz $\sigma^{ab}_D = \frac{1}{2}S^{ab}C_D$ for the spin de Rham current $\sigma^{ab}_D$ gives 
\begin{align}
\nabla_{\dot{C}}S^{ab}&= -\dot{C}^a\left({\pi}^b+(i_{X^b}{F})\cdot i_{\dot{C}}\Sigma\right)-(i_{X^b}{F})\cdot i_{X^a}\Sigma\nonumber\\
  &\quad+\dot{C}^b\left({\pi}^a+(i_{X^a}{F})\cdot i_{\dot{C}}\Sigma\right)+(i_{X^a}{F})\cdot i_{X^b}\Sigma.\label{chieqofmotion1}
\end{align}

\subsection{Substituting the Nakano-Tulczyjew momentum into the equations of motion}

Since the spin matching condition commonly used is the Nakano-Tulczyjew condition \eqref{Tulczcond}, it is logical to write the equations of motion \eqref{Peqofmotion1} and \eqref{chieqofmotion1} in terms of the 4-momentum $P^a$ of the particle. This momentum  may be expressed as
\begin{align}
P^a(\lambda) = -\int_{\Sigma_\lambda}T^{ab}N_b\star\widetilde{N},
\end{align}
where $T^{ab}$ is the stress-energy-momentum tensor and the 1-parameter family $\Sigma_\lambda$ of spacelike hypersurfaces is the set of leaves of a local foliation of spacetime with timelike unit normal $N=\frac{P}{|P|}$. Since the stress-energy-momentum tensor $T^{ab}$ is related to the stress-energy-momentum forms $\mathcal{T}^a$ via $\mathcal{T}^a = \star\left(T(X^a,-)\right)$, note that for test 0-form $\hat{f}$,
\begin{align}
\int_{\mathcal{M}}\mathcal{T}^a\wedge\widetilde{N}\hat{f} = \int_{\mathcal{M}}\star\left(T(X^a,-)\right)\wedge\widetilde{N}\hat{f}.
\end{align}
Utilising the identity $\star\alpha\wedge\widetilde{N} = \star\widetilde{N}\wedge\alpha$ where $\alpha$ is a 1-form  and noting that the vector $N$ is normalised as $g(N,N)=-1$, so that the volume form can be written $\star 1 = -\widetilde{N}\wedge \star \widetilde{N}$, this can be simplified to
\begin{align}
\int_{\mathcal{M}}\mathcal{T}^a\wedge\widetilde{N}\hat{f} = \int_{\mathcal{M}}\left(T(X^a,N)\right)\widetilde{N}\wedge \star \widetilde{N}\hat{f}.
\end{align}

Since $\widetilde{N}=-\frac{d\lambda}{|d\lambda|}$, the integral can be split into a piece along the worldline $C$ and another over the hyperplane $\Sigma_\lambda$ via
\begin{align}
\int_{\mathcal{M}}\mathcal{T}^a\wedge\widetilde{N} \hat{f} &= -\int_C\frac{d\lambda}{|d\lambda|}\int_{\Sigma_\lambda}T^{ab}N_b\star \widetilde{N}\hat{f}\nonumber\\
&=-\int_CP^a\frac{d\lambda}{|d\lambda|}\hat{f},
\end{align}
and stripping off the test forms yields the relation
\begin{align}
\mathcal{T}^a_{D}\wedge\widetilde{N} = -P^aC_D\wedge\widetilde{N},
\end{align}
with $N^a = \frac{P^a}{|P|}$. Using the stress forms $\mathcal{T}^a_{D}$ given by \eqref{taukeq} yields 
\begin{align}
\left(-g(\mathbf{\pi},{X^a})C_D + i_{\dot{C}}\left(i_{X^a}{F}\wedge\star\Sigma\right)\wedge\star C_D\right)\wedge \widetilde{P} &= -P^aC_D\wedge \widetilde{P},
\end{align}
which upon manipulation yields the relationship
\begin{align}
\pi^a  = P^a - F^{ab}\Sigma_{cb}\left(\frac{{P}^c}{\left(P\cdot {\dot{C}}\right)} + \dot{C}^c\right)\label{Ppitaurelationproofed},
\end{align}
between the momenta $\pi^a$ and $P^a$. Substitution of \eqref{Ppitaurelationproofed} into the equations of motion \eqref{Peqofmotion1} and \eqref{chieqofmotion1} yields the familiar equations 
\begin{align}
&\frac{d}{d\tau}\left(P^a + \frac{{F}^{ab}\Sigma_{bc} {P}^c}{\dot{C}^dP_d}\right)= -q{F}^{ab}{\dot{C}_b}-\frac{1}{2}\Sigma^{bc}\partial^a{F}_{bc},\label{SDGEQ1}\\
&\frac{d}{d\tau}S^{ab}= -\dot{C}^a\left(P^b + \frac{{F}^{bc}\Sigma_{cd} {P}^d}{\dot{C}^eP_e}\right)+\dot{C}^b\left(P^a + \frac{{F}^{ac}\Sigma_{cd} {P}^d}{\dot{C}^eP_e}\right) +{F}^{bc}\Sigma^{~a}_{c} -{F}^{ac}\Sigma^{~b}_{c}\label{SDGEQ2}
\end{align}
found in the literature \cite{suttorpdegroot}.

\section{Plasma wave electric field}
\label{App:PW}
A sufficiently short and intense laser pulse propagating through a plasma drives a non-linear wave in the electron number density. For present purposes, the ions are essentially stationary over the timescales of interest because their charge to mass ratio is about three orders of magnitude lower than that of the electrons.

Despite the recent focus on the three dimensional `bubble regime' \cite{Luetal2006,Yietal2013}, one-dimensional models remain useful for providing estimates, particularly in contexts such as the Stern-Gerlach force in the main body of the paper, where additional complexity leads to dramatically more difficult analysis. Some preliminary work on the subject of electrons in a one-dimensional maximum amplitude plasma wave has already been done in the context of non-linear electrodynamics \cite{DABPlasmas}, and we use the same expressions for the plasma electron worldlines and overall electric field; a brief summary follows.

Assuming the electric field is due only to the electron fluid and the ion background, the magnetic field vanishes, leaving only the electric field component in the direction of the propagation of the wave. The Faraday 2-form of such a  wave driven by a laser pulse in the $z$-direction is thus simply
\begin{align}
F = E(\xi)dt\wedge dz\label{PWFdef},
\end{align}
where $\xi = z-vt$ is the phase of the wave. The electric field $E$ is governed by the field and Lorentz force equations for a cold plasma:
\begin{align}
dF &= 0\label{Maxwell1plasma},\\
d\star F &= -q_{\text{e}}n_{\text{e}}\star \widetilde{V}_{\text{e}} - q_{\text{ion}}n_{\text{ion}}\star \widetilde{V}_{\text{ion}},\label{Maxwell2plasma}\\
\nabla_{V_{\text{e}}}\widetilde{V}_{\text{e}} &= \frac{q_{\text{e}}}{m_{\text{e}}}i_{V_{\text{e}}}F,\label{PWelectronlorF}
\end{align}
where $q_{\text{ion}}$ is the ion charge, $n_{\text{e}}$ and $n_{\text{ion}}$ are the proper number densities of the electrons and ions, $V_{\text{e}}$ and $V_{\text{ion}}$ are the 4-vectors whose trajectories are the worldlines of the electrons and ions, respectively. For simplicity, the plasma electrons are assumed to be unpolarised and so the motion of the electrons is governed purely by the usual Lorentz force.

For a plasma wave whose electron motion is much greater than the motion of the ions, it is reasonable to assume that the plasma ions are at rest, i.e. $V_{\text{ion}} = \partial_t$ with constant number density $n_{\text{ion}}$. A simple example is laser-plasma wakefield acceleration, where the plasma wave electrons oscillate far faster than the motion of the plasma ions. Seeking a 4-vector describing the motion of the electrons of the form
\begin{align}
\widetilde{V_{\text{e}}} = \nu(\xi)d\zeta - \psi(\xi)d\xi,
\end{align}
the normalisation condition $g(V_{\text{e}},V_{\text{e}})=-1$ restricts $\psi$, giving
\begin{align}
\widetilde{V_{\text{e}}} = \nu(\xi)d\zeta - \sqrt{\nu(\xi)^2-\gamma^2}d\xi,\label{PWelectronworldlinedef}
\end{align}
where the negative sign is chosen so that the electrons move at speed less than the wave (for more details, see \cite{DABPlasmas}). In order to satisfy the Lorentz force \eqref{PWelectronlorF}, the electric field must be of the form
\begin{align}
E = \frac{m_{\text{e}}\nu'}{q_{\text{e}}\gamma^2},\label{PWEintermsofnudef2}
\end{align}
so that the function $\nu$ is similar to the electric potential.  Similarly in order to satisfy the field equations \eqref{Maxwell1plasma} and \eqref{Maxwell2plasma} the electron number density can be written in terms of the ion density:
\begin{align}
n_{\text{e}} = -\frac{v\gamma^2q_{\text{ion}}n_{\text{ion}}}{q_{\text{e}}\sqrt{\nu^2-\gamma^2}}=  \frac{v\gamma^2Zn_{\text{ion}}}{\sqrt{\nu^2-\gamma^2}},
\end{align}
where $Z$ is the degree of ionisation, $Z=-\frac{q_{\text{ion}}}{q_{\text{e}}}$, and $\nu$ must satisfy
\begin{align}
\frac{d}{d\xi}\left[\frac{m_{\text{e}}^2}{2q_{\text{e}}^2\gamma^4}\nu'^2 - m_{\text{e}}Zn_{\text{ion}}\left(v\sqrt{\nu^2-\gamma^2}-\nu\right)\right] = 0.\label{smallernucondition2inprogress}
\end{align}
The square root in \eqref{smallernucondition2inprogress} imposes a lower bound on $\nu$. This results in a maximum amplitude oscillation for the plasma wave, and the maximum amplitude is known as the wave-breaking limit \cite{dawson1959nonlinear}. The  wave-breaking limit may be obtained from \eqref{smallernucondition2inprogress} by integrating from $\xi_{\text{I}}$, the minimum of $\nu$ and hence a zero of $E$, to $\xi_{\text{II}}$, the maximum of $E$ and turning point of $\nu'$. Since $\nu_{\text{I}} = \nu(\xi_{\text{I}}) = \gamma$  and $\nu_{\text{II}}=\gamma^2$ (from \eqref{smallernucondition2inprogress}), it follows that 
\begin{align}
E_{\text{\text{max}}} = \sqrt{2(\gamma-1)m_{\text{e}}Zn_{\text{ion}}}.
\end{align}
Given that the plasma wave attains its lowest possible value at $\xi_{\text{I}}$, integrating \eqref{smallernucondition2inprogress} between two subsequent zeros of the electric field $\xi_{\text{I}}$ and $\xi_{\text{III}}$ where $\nu(\xi_{\text{III}}) = \gamma^3(1+v^2)$ gives 
\begin{align}
\frac{m_{\text{e}}^2}{2q_{\text{e}}^2\gamma^4}\nu'^2 - m_{\text{e}}Zn_{\text{ion}}\left(v\sqrt{\nu^2-\gamma^2}-\nu+\gamma\right) = 0,\label{smallernucondition2}
\end{align}
governing the electric field in the case of a maximum amplitude oscillation. 

\bibliography{refs}{}
\bibliographystyle{ieeetr}

\end{document}